\documentstyle[prd,aps,preprint,tighten,epsfig]{revtex}

\begin{document}

\draft
\preprint{\begin{tabular}{r}
\framebox{\bf hep-ph/0107005} \\
{~}
\end{tabular}}

\title{Almost Maximal Lepton Mixing with Large T Violation in
Neutrino Oscillations and Neutrinoless Double Beta Decay}
\author{\bf Zhi-zhong Xing}
\address{Institute of High Energy Physics, P.O. Box 918 (4), 
Beijing 100039, China \\
(Electronic address: xingzz@mail.ihep.ac.cn)}
\maketitle

\begin{abstract}
We point out two simple but instructive possibilities to construct
the charged lepton and neutrino mass matrices, from which the nearly 
bi-maximal neutrino mixing with large T violation can naturally emerge.
The two lepton mixing scenarios are compatible very well with
current experimental data on solar and atmospheric neutrino oscillations,
and one of them may lead to an observable T-violating asymmetry between 
$\nu_\mu \rightarrow \nu_e$ and $\nu_e \rightarrow \nu_\mu$ transitions 
in the long-baseline neutrino oscillation experiments. Their implications
on the neutrinoless double beta decay are also discussed.  
\end{abstract}


\section{Introduction}

Recent observation of atmospheric and solar neutrino anomalies, particularly
that in the Super-Kamiokande experiment \cite{SK}, has provided 
robust evidence that neutrinos are massive and lepton flavors are mixed.
Analyses of the atmospheric neutrino deficit favor 
$\nu_\mu \rightarrow \nu_\tau$ as the dominant oscillation 
mode with the mass-squared difference 
$\Delta m^2_{\rm atm} \sim 10^{-3} ~ {\rm eV^2}$ and the
mixing factor $\sin^2 2\theta_{\rm atm} > 0.88$ at the
$90\%$ confidence level.
As for the solar neutrino anomaly,
there are four possible solutions belonging to 
two categories:
(a) solar $\nu_e$ neutrinos changing to active $\nu_\mu$ or 
sterile $\nu_s$ neutrinos due to the long-wavelength vacuum 
oscillation with the parameters
$\Delta m^2_{\rm sun} \sim 10^{-10} ~ {\rm eV^2}$ and
$\sin^2 2\theta_{\rm sun} \approx 1$ \cite{VO};
(b) the matter-enhanced $\nu_e\rightarrow \nu_\mu$ or
$\nu_e\rightarrow \nu_s$ oscillations via the 
Mikheyev-Smirnov-Wolfenstein (MSW) mechanism with 
$\Delta m^2_{\rm sun} \sim 10^{-5} ~ {\rm eV^2}$ and
$\sin^2 2\theta_{\rm sun} \sim 1$ (large-angle solution),
with $\Delta m^2_{\rm sun} \sim 10^{-6} ~ {\rm eV}^2$ and
$\sin^2 2\theta_{\rm sun} \sim 10^{-2}$ (small-angle solution),
or with $\Delta m^2_{\rm sun} \sim 10^{-7} ~ {\rm eV}^2$ and
$\sin^2 2\theta_{\rm sun} \sim 1$ (low solution) \cite{Bahcall}. 
Although the large-angle MSW solution seems to be somehow favored by the
present Super-Kamiokande and SNO data \cite{SK,SNO}, 
the other three solutions have not been convincingly ruled out.
To pin down the true solution to the solar neutrino problem 
remains a challenging task of the 
next-round solar neutrino experiments.

The strong hierarchy between 
$\Delta m^2_{\rm atm}$ and $\Delta m^2_{\rm sun}$,
together with the small $\nu_3$-component in $\nu_e$ 
configuration \cite{CHOOZ},
implies that atmospheric and solar neutrino oscillations decouple 
approximately from each other. Each of them is 
dominated by a single mass scale
\footnote{Throughout this paper we do not take the LSND evidence for
neutrino oscillations \cite{LSND}, which has not been independently
confirmed by other experiments \cite{KARMEN}, into account.}, 
which can be set as
\begin{eqnarray}
\Delta m^2_{21} & \equiv &
m^2_2 - m^2_1 \; \approx \; \pm \Delta m^2_{\rm sun} \; ,
\nonumber \\
\Delta m^2_{32} & \equiv &
m^2_3 - m^2_2 \; \approx \; \pm \Delta m^2_{\rm atm} \; .
\end{eqnarray}
Of course $\Delta m^2_{31} \approx \Delta m^2_{32}$ holds in this
approximation. As a consequence,
the mixing factors of solar and atmospheric
neutrino oscillations in the disappearance-type experiments
(i.e., $\nu_e \rightarrow \nu_e$ and $\nu_\mu \rightarrow \nu_\mu$)
are simply given by
\begin{eqnarray}
\sin^2 2\theta_{\rm sun} & = & 4 |V_{e1}|^2 |V_{e2}|^2 \; ,
\nonumber \\
\sin^2 2\theta_{\rm atm} & = & 4 |V_{\mu 3}|^2 \left ( 1- 
|V_{\mu 3}|^2 \right ) \; ,
\end{eqnarray}
where $V$ is the $3\times 3$ lepton flavor mixing matrix linking
the neutrino mass eigenstates $(\nu_1, \nu_2, \nu_3)$
to the neutrino flavor eigenstates $(\nu_e, \nu_\mu, \nu_\tau)$.
The present experimental data seem to favor the large-angle
MSW solution to the solar neutrino problem. In this case,
$\sin^2 2\theta_{\rm atm} \sim \sin^2 2\theta_{\rm sun} 
\sim {\cal O}(1)$. Then two large mixing angles can be drawn from
Eq. (2): one between the 2nd and 3rd lepton families and 
the other between the 1st and 2nd lepton families
\footnote{
The conjecture, that two of the three lepton flavor mixing angles could 
be extraordinarily large (i.e., equal or close to $45^\circ$)
had been made by several authors \cite{FX96}
before the first-round Super-Kamiokande data appeared in 1998.}.

A particularly interesting limit is 
$\sin^2 2\theta_{\rm atm} = \sin^2 2\theta_{\rm sun} =1$,
corresponding uniquely (up to a trivial sign or phase rearrangement) to 
\begin{equation}
V_0 \; =\; \left ( \matrix{
\frac{1}{\sqrt{2}}      & \frac{1}{\sqrt{2}}    & 0 \cr\cr
-\frac{1}{2}    & \frac{1}{2}   & \frac{1}{\sqrt{2}} \cr\cr
\frac{1}{2}     & -\frac{1}{2}  & \frac{1}{\sqrt{2}} \cr} \right ) \;\; ,
\end{equation}
the so-called ``bi-maximal'' flavor mixing pattern \cite{Bi}.
There have been a lot of discussions about 
the bi-maximal and nearly bi-maximal neutrino mixing scenarios \cite{FX00}.
While the latter could straightforwardly be obtained from slight
modifications of the former, the arbitrariness in doing so 
has to be resolved by imposing simple
flavor symmetries or dynamical constraints 
on the charged lepton and neutrino mass matrices.
In Ref. \cite{FX98}, for example, it has been shown that a nearly bi-maximal 
neutrino mixing pattern can naturally arise from the explicit breaking 
of the lepton flavor democracy. 

The present paper aims to discuss two simple but 
instructive possibilities to construct the lepton mass
matrices, from which two almost bi-maximal neutrino mixing patterns  
can directly be derived. We find that these two scenarios have
practically indistinguishable consequences on solar and
atmospheric neutrino oscillations, but their predictions for
leptonic CP or T violation are quite different.
To be specific, we calculate the deviation of solar neutrino
mixing from maximal mixing in each scenario. We illustrate
that one of the two lepton mixing patterns may lead to an observable 
T-violating asymmetry between $\nu_\mu \rightarrow \nu_e$ and
$\nu_e \rightarrow \nu_\mu$ transitions in the long-baseline
neutrino oscillation experiments. The implications of our phenomenological
models on the neutrinoless double beta decay are also discussed in some
detail.

\section{Nearly bi-maximal mixing}

The fact that the masses of three active neutrinos are extremely small
is presumably attributed
to the Majorana feature of the neutrino fields \cite{Seesaw}. 
In this picture, the light (left-handed) neutrino mass matrix 
$M_\nu$ must be symmetric and can be diagonalized by a single
unitary transformation:
\begin{equation}
U^\dagger_\nu M_\nu U^*_\nu \; = \; \left ( \matrix{
m_1 & 0 & 0 \cr
0 & m_2 & 0 \cr
0 & 0 & m_3 \cr} \right ) \; .
\end{equation}
The charged lepton mass
matrix $M_l$ is in general non-Hermitian, hence the
diagonalization of $M_l$ needs a bi-unitary transformation:
\begin{equation}
U^{\dagger}_l M_l \; \tilde{U}_l \; = \; \left ( \matrix{
m_e  & 0   & 0 \cr
0 & m_\mu  & 0 \cr
0 & 0 & m_\tau \cr} \right ) \; . 
\end{equation}
The lepton flavor mixing matrix $V$, defined to link the
neutrino mass eigenstates $(\nu_1, \nu_2, \nu_3)$ to the
neutrino flavor eigenstates $(\nu_e, \nu_\mu, \nu_\tau)$,
measures the
mismatch between the diagonalization of $M_l$ and that of $M_\nu$:
\begin{equation}
V \; =\; U^{\dagger}_l U_\nu \; .
\end{equation}
Note that $(m_1, m_2, m_3)$ in Eq. (4) and 
$(m_e, m_\mu, m_\tau)$ in Eq. (5) are physical (positive) masses
of light neutrinos and charged leptons, respectively.

In the flavor basis where
$M_l$ is diagonal (i.e., $U_l = {\bf 1}$ being a unity matrix),
the flavor mixing matrix is simplified to $V = U_\nu$. The bi-maximal
neutrino mixing pattern $U_\nu =V_0$ can then be constructed from the
product of the Euler rotation matrices 
\begin{equation}
R_{12}(\theta_x) \; =\; \left ( \matrix{
\cos\theta_x	& \sin\theta_x 	& 0 \cr
-\sin\theta_x	& \cos\theta_x	& 0 \cr
0	& 0	& 1 \cr} \right ) \;
\end{equation}
and 
\begin{equation}
R_{23}(\theta_y) \; = \; \left ( \matrix{
1	& 0	& 0 \cr
0	& \cos\theta_y	& \sin\theta_y \cr
0	& -\sin\theta_y	& \cos\theta_y \cr} \right )
\end{equation}
with special rotation angles $\theta_x = \theta_y = 45^\circ$:
\begin{eqnarray}
V_0 & = & R_{23}(45^\circ) \otimes R_{12}(45^\circ) 
\nonumber \\ 
& = & \left ( \matrix{
\frac{1}{\sqrt{2}}      & \frac{1}{\sqrt{2}}    & 0 \cr\cr
-\frac{1}{2}    & \frac{1}{2}   & \frac{1}{\sqrt{2}} \cr\cr
\frac{1}{2}     & -\frac{1}{2}  & \frac{1}{\sqrt{2}} \cr} \right ) \;\; .
\end{eqnarray}
Obviously the vanishing of the (1,3) element in $V_0$ assures
an exact decoupling between solar ($\nu_e \rightarrow \nu_\mu$) 
and atmospheric ($\nu_\mu \rightarrow \nu_\tau$) neutrino oscillations.
The corresponding neutrino mass matrix $M_\nu$ turns out to be
\begin{eqnarray}
M_\nu & = & V_0 \left ( \matrix{
m_1 & ~~ 0 ~~ & 0 \cr
0 & ~~ m_2 ~~ & 0 \cr
0 & ~~ 0 ~~ & m_3 \cr} \right ) V^{\rm T}_0
\nonumber \\ 
& = & \left ( \matrix{
A_\nu - B_\nu & C_\nu & -C_\nu \cr
C_\nu & A_\nu & B_\nu \cr
-C_\nu & B_\nu & A_\nu \cr} \right ) \; ,
\end{eqnarray}
where 
\begin{eqnarray}
A_\nu & = & \frac{m_3}{2} ~ + ~ \frac{m_1 + m_2}{4}  \;\; , 
\nonumber \\
B_\nu & = & \frac{m_3}{2} ~ - ~ \frac{m_1 + m_2}{4}  \;\; , 
\nonumber \\
C_\nu & = & \frac{m_2 - m_1}{2\sqrt{2}} \;\; .
\end{eqnarray}
A trivial sign or phase rearrangement for $U_\nu =V_0$ may lead to
a slightly different form of $M_\nu$ \cite{Bi,Mohapatra}, 
but the relevant physical consequences on neutrino oscillations
are essentially unchanged.
If the masses of $\nu_1$ and $\nu_2$ neutrinos
are nearly degenerate (i.e., $m_1 \approx m_2$), one can arrive at 
a simpler texture of $M_\nu$, in which $A_\nu \approx (m_3 + m_1)/2$, 
$B_\nu \approx (m_3 - m_1)/2$, and $C_\nu \approx 0$ hold.

We observe that the bi-maximal neutrino mixing pattern will be
modified, if $U_l$ deviates somehow from the unity matrix. This
can certainly happen, provided that the charged lepton mass matrix $M_l$ is
not diagonal in the flavor basis where the neutrino mass matrix
$M_\nu$ takes the form given in Eq. (10). 
As $U_\nu = V_0$ describes a product of 
two special Euler rotations in the real (2,3) and
(1,2) planes, the simplest form of $U_l$ which allows $V=U^\dagger_l U_\nu$ 
to cover the whole $3\times 3$ space should be 
$U_l = R_{12}(\theta_x)$ or $U_l = R_{31}(\theta_z)$ 
(see Ref. \cite{Xing98} for a detailed discussion). 
To incorporate T violation in neutrino oscillations, however, 
the complex rotation matrices 
\begin{equation}
R_{12}(\theta_x, \phi_x) \; =\; \left ( \matrix{
\cos\theta_x	& \sin\theta_x e^{i\phi_x}	& 0 \cr
-\sin\theta_x e^{-i\phi_x}	& \cos\theta_x	& 0 \cr
0	& 0	& 1 \cr} \right ) \;
\end{equation}
and
\begin{equation}
R_{31}(\theta_z, \phi_z) \; =\; \left ( \matrix{
\cos\theta_z	& 0	& \sin\theta_z e^{i\phi_z} \cr
0	& 1	& 0 \cr
-\sin\theta_z e^{-i\phi_z}	& 0	& \cos\theta_z	\cr} \right )
\end{equation}
should be used \cite{FX01}. In this case, we arrive at
lepton flavor mixing of the pattern 
\begin{eqnarray}
V_{(x)} & = & \left ( \matrix{
c_x  & -s_x e^{i\phi_x}    & ~ 0 \cr
s_x e^{-i\phi_x}  & c_x     & ~ 0 \cr
0           & 0              & ~ 1 \cr} \right ) 
\left ( \matrix{
\frac{1}{\sqrt{2}}   & ~ \frac{1}{\sqrt{2}} ~  & 0 \cr
-\frac{1}{2}   & ~ \frac{1}{2} ~ & \frac{1}{\sqrt{2}} \cr
\frac{1}{2}    & ~ -\frac{1}{2} ~ & \frac{1}{\sqrt{2}} \cr} \right )
\nonumber \\ \nonumber \\ 
& = & \left ( \matrix{
\frac{\displaystyle c_x}{\sqrt{2}} + \frac{\displaystyle
s_x}{2} e^{i\phi_x}  ~ & 
\frac{\displaystyle c_x}{\sqrt{2}} - \frac{\displaystyle 
s_x}{2} e^{i\phi_x} ~ & 
-\frac{\displaystyle s_x}{\sqrt{2}} e^{i\phi_x} \cr\cr
-\frac{\displaystyle c_x}{2} + \frac{\displaystyle
s_x}{\sqrt{2}} e^{-i\phi_x} ~ & 
\frac{\displaystyle c_x}{2} + \frac{\displaystyle 
s_x}{\sqrt{2}} e^{-i\phi_x} ~ &
\frac{\displaystyle c_x}{\sqrt{2}} \cr\cr
\frac{1}{2}  & -\frac{1}{2}   & \frac{1}{\sqrt{2}} \cr} \right ) \;\; ,
\end{eqnarray}
or of the pattern
\begin{eqnarray}
V_{(z)} & = & \left ( \matrix{
c_z  & ~ 0 ~    & -s_z e^{i\phi_z} \cr
0  & ~ 1 ~     & 0 \cr
s_z e^{-i\phi_z}  & ~ 0 ~    & c_z \cr} \right ) 
\left ( \matrix{
\frac{1}{\sqrt{2}}   & ~ \frac{1}{\sqrt{2}} ~  & 0 \cr
-\frac{1}{2}   & ~ \frac{1}{2} ~  & \frac{1}{\sqrt{2}} \cr
\frac{1}{2}    & ~ -\frac{1}{2} ~ & \frac{1}{\sqrt{2}} \cr} \right )
\nonumber \\ \nonumber \\ 
& = & \left ( \matrix{ 
\frac{\displaystyle c_z}{\sqrt{2}} - \frac{\displaystyle
s_z}{2} e^{i\phi_z} ~ & 
\frac{\displaystyle c_z}{\sqrt{2}} + \frac{\displaystyle 
s_z}{2} e^{i\phi_z} ~ & 
-\frac{\displaystyle s_z}{\sqrt{2}} e^{i\phi_z} \cr\cr
-\frac{1}{2}    & \frac{1}{2}    & \frac{1}{\sqrt{2}} \cr\cr
\frac{\displaystyle c_z}{2} + \frac{\displaystyle
s_z}{\sqrt{2}} e^{-i\phi_z} ~ & 
-\frac{\displaystyle c_z}{2} + \frac{\displaystyle 
s_z}{\sqrt{2}} e^{-i\phi_z} ~ &
\frac{\displaystyle c_z}{\sqrt{2}} \cr} \right ) \;\; ,
\end{eqnarray}
where $s_x \equiv \sin \theta_x$, $c_z \equiv \cos\theta_z$,
and so on. 
It is obvious that $V_{(x)}$ and $V_{(z)}$ represent two 
nearly bi-maximal flavor mixing scenarios, if the rotation 
angles $\theta_x$ and $\theta_z$ are small in magnitude,

As the mixing angle $\theta_x$ or $\theta_z$ arises from
the diagonalization of $M_l$, it is expected to be a simple function
of the ratios of charged lepton masses. Then the strong mass
hierarchy of charged leptons naturally assures the
smallness of $\theta_x$ or $\theta_z$, as one can see later on.

\section{Constraints on $\sin^2 2\theta_{\rm sun}$ and
$\sin^2 2\theta_{\rm atm}$}

Indeed the proper texture of $M_l$ 
which leads to the flavor mixing pattern $V_{(x)}$ is 
\begin{equation}
M^{(x)}_l \; =\; \left ( \matrix{
0  & C_l & 0 \cr
C^*_l & B_l  & 0 \cr
0 & 0 & A_l \cr} \right ) \;\; ,
\end{equation}
where $A_l = m_\tau$, $B_l = m_\mu - m_e$, and 
$C_l =\sqrt{m_e m_\mu} ~ e^{i\phi_x}$.
The mixing angle $\theta_x$ in $V_{(x)}$ is then given by
\begin{equation}
\tan (2\theta_x) \;\; =\;\; 2 ~ \frac{\sqrt{m_e m_\mu}}{m_\mu - m_e} \;\; .
\end{equation}
On the other hand, the proper texture of $M_l$ which gives rise to
the mixing pattern $V_{(z)}$ reads as follows: 
\begin{equation}
M^{(z)}_l \; =\; \left ( \matrix{
0  & 0 & C_l \cr
0  & B_l  & 0 \cr
C^*_l & 0 & A_l \cr} \right ) \;\; ,
\end{equation}
where $A_l = m_\tau -m_e$, $B_l = m_\mu$, and 
$C_l =\sqrt{m_e m_\tau} ~ e^{i\phi_z}$.
The mixing angle $\theta_z$ in $V_{(z)}$ turns out to be
\begin{equation}
\tan (2\theta_z) \;\; =\;\; 2 ~ \frac{\sqrt{m_e m_\tau}}{m_\tau - m_e} \;\; .
\end{equation}
Taking the hierarchy of charged lepton masses 
(i.e., $m_e \ll m_\mu \ll m_\tau$) into account,
one obtains 
\begin{eqnarray}
s_x & \approx & \sqrt{\frac{m_e}{m_\mu}} \;\; ,
\nonumber \\
s_z & \approx & \sqrt{\frac{m_e}{m_\tau}} \;\; ,
\end{eqnarray}
to a good degree of accuracy.
Numerically, we find $\theta_x \approx 3.978^\circ$ and
$\theta_z \approx 0.972^\circ$ with the inputs 
$m_e = 0.511$ MeV, $m_\mu = 105.658$ MeV, and
$m_\tau = 1.777$ GeV \cite{PDG}. 

Now let us calculate the mixing factors of solar and
atmospheric neutrino oscillations in the disappearance-type experiments.
Using Eq. (2), we arrive straightforwardly at
\begin{eqnarray}
\sin^2 2\theta_{\rm sun} & = & 1 - s^2_x \left ( 1+
2 \cos^2 \phi_x \right ) \; ,
\nonumber \\
\sin^2 2\theta_{\rm atm} & = & 1 - s^4_x \; 
\end{eqnarray}
for $V_{(x)}$; and 
\begin{eqnarray}
\sin^2 2\theta_{\rm sun} & = & 1 - s^2_z \left ( 1+
2 \cos^2 \phi_z \right ) \; ,
\nonumber \\
\sin^2 2\theta_{\rm atm} & = & 1 \; 
\end{eqnarray}
for $V_{(z)}$. Allowing $\phi_x$ and $\phi_z$ to take 
arbitrary values, we find that the magnitude of 
$\sin^2 2\theta_{\rm sun}$ lies in the following range:
\begin{equation}
1 - 3s^2_i \; \leq \; \sin^2 2\theta_{\rm sun} \;
\leq \; 1 - s^2_i \; ,
\end{equation}
where $i=x$ or $z$.
Numerically, we obtain $0.986 \leq \sin^2 2\theta_{\rm sun} \leq 0.995$
for $V_{(x)}$ and $0.999 \leq \sin^2 2\theta_{\rm sun} \leq 1.000$
for $V_{(z)}$. Note that $\sin^2 2\theta_{\rm atm} = 1.000$ holds in
both cases.
Therefore the two nearly bi-maximal neutrino mixing patterns
are practically indistinguishable in the experiments
of solar and atmospheric neutrino oscillations. 
They may be distinguished from each other with the measurements 
of $|V_{e3}|$ and CP or T violation in the long-baseline
neutrino oscillation experiments. 

It is worth mentioning that 
Gonzalez-Garcia, Pe$\rm\tilde{n}$a-Garay,
Nir, and Smirnov have recently defined a small real parameter $\epsilon$
to describe the deviation of solar neutrino mixing from maximal
mixing \cite{Smirnov}:
\begin{equation}
\sin^2 \theta_{\rm sun} \; \equiv \; \frac{1 - \epsilon}{2} 
\end{equation}
with $|\epsilon| \ll 1$. This parameter proves very useful for 
phenomenological studies of the solar neutrino problem \cite{Smirnov}:
the probabilities of solar neutrino oscillations depend quadratically 
on $\epsilon$ in vacuum, and linearly on $\epsilon$ if matter
effects dominate. It is then our interest to calculate $\epsilon$ 
in the nearly bi-maximal neutrino mixing scenarios under discussion.
We notice that
\begin{equation}
\sin^2 2 \theta_{\rm sun} \; =\; 1 - \epsilon^2 \; 
\end{equation}
results from Eq. (24) exactly. Comparing Eq. (25) with
Eqs. (21) and (22), we obtain
\begin{equation}
|\epsilon| \; =\; s_x \sqrt{1 + 2 \cos^2\phi_x} \; 
\end{equation}
for $V_{(x)}$; and
\begin{equation}
|\epsilon| \; =\; s_z \sqrt{1 + 2 \cos^2\phi_z} \; 
\end{equation}
for $V_{(z)}$. Given $\phi_x$ and $\phi_z$ of 
arbitrary values, the allowed region of $|\epsilon|$ turns out to
be $0.069 \leq |\epsilon| \leq 0.120$ in the scenario of
$V_{(x)}$, and $0.017 \leq |\epsilon| \leq 0.029$ in the scenario
of $V_{(z)}$. Both ranges of $|\epsilon|$ are phenomenologically
interesting for solar neutrino oscillations, as comprehensively
discussed in Ref. \cite{Smirnov}. 

\section{Leptonic T violation}

The strength of CP or T violation in neutrino oscillations,
no matter whether neutrinos are Dirac or Majorana particles,
is measured by a universal and rephasing-invariant parameter
$\cal J$ \cite{Xing98}, defined through the following equation:
\begin{equation}
{\rm Im} \left (V_{\alpha i} V_{\beta j} V^*_{\alpha j} V^*_{\beta i} \right ) 
\; =\; {\cal J} \sum_{\gamma, k} \left (\varepsilon_{\alpha\beta\gamma}
\varepsilon_{ijk} \right ) \; ,
\end{equation}
in which the Greek subscripts run over $(e, \mu, \tau)$, and the 
Latin subscripts run over $(1,2,3)$.
Considering the two lepton mixing scenarios proposed
in section 2, we obtain 
\begin{equation}
{\cal J} \; = \; \left \{ \matrix{
\displaystyle \frac{c_x s_x}{4\sqrt{2}} \sin \phi_x
~~~~~~~~ [{\rm for} ~ V_{(x)}] \; ,  \cr\cr\cr
\displaystyle \frac{c_z s_z}{4\sqrt{2}} \sin \phi_z 
~~~~~~~~ [{\rm for} ~ V_{(z)}] \; . \cr} \right . 
\end{equation}
For illustration,
we typically take $\phi_x = \phi_z = 90^\circ$. Then
we arrive at ${\cal J} \approx 0.012$ and
${\cal J} \approx 0.003$, respectively, for $V_{(x)}$ and $V_{(z)}$. 
The former could be determined from the probability asymmetry 
between $\nu_\mu \rightarrow \nu_e$ and 
$\bar{\nu}_\mu \rightarrow \bar{\nu}_e$ transitions 
(CP-violating asymmetry), or that
between $\nu_\mu \rightarrow \nu_e$ and 
$\nu_e \rightarrow \nu_\mu$ transitions (T-violating asymmetry)
in a long-baseline neutrino oscillation experiment \cite{Xing01}, if the
earth-induced matter effects were assumed to be absent or negligible:
\begin{eqnarray}
\Delta P & = & P (\nu_\mu \rightarrow \nu_e)
- P (\bar{\nu}_\mu \rightarrow \bar{\nu}_e) 
\nonumber \\
& = & P (\nu_\mu \rightarrow \nu_e)
- P (\nu_e \rightarrow \nu_\mu) 
\nonumber \\
& = & 16 ~ {\cal J} \sin F_{12} \sin F_{23} \sin F_{31} 
\nonumber \\
& \approx & 16 ~ {\cal J} \sin F_{21} \sin^2 F_{32} \; ,
\end{eqnarray}
where $F_{ij} = 1.27 \Delta m^2_{ij} L/E$ with $L$ being the distance
between the neutrino source and the detector (in unit of km) and 
$E$ being the neutrino beam energy (in unit of GeV).
In realistic long-baseline neutrino oscillation experiments,
however, the terrestrial matter effects are by no means 
small and must be taken into account. 

It is generally expected that the T-violating asymmetry between
$\nu_\mu \rightarrow \nu_e$ and $\nu_e \rightarrow \nu_\mu$ transitions
is less sensitive to matter effects than the CP-violating asymmetry
between $\nu_\mu \rightarrow \nu_e$ and 
$\bar{\nu}_\mu \rightarrow \bar{\nu}_e$ transitions \cite{Kuo}. For
simplicity, we concentrate only on T violation in the following.
In analogy to Eq. (30),
the matter-corrected T-violating asymmetry can be expressed as
\begin{eqnarray}
\Delta \tilde{P} & = & \tilde{P} (\nu_\mu \rightarrow \nu_e)
- \tilde{P} (\nu_e \rightarrow \nu_\mu) 
\nonumber \\
& = & 16 ~ \tilde{\cal J} \sin \tilde{F}_{12} 
\sin \tilde{F}_{23} \sin \tilde{F}_{31} 
\nonumber \\
& \approx & 16 ~ \tilde{\cal J} \sin \tilde{F}_{21} 
\sin^2 \tilde{F}_{32} \; ,
\end{eqnarray}
where $\tilde{F}_{ij} = 1.27 \Delta \tilde{m}^2_{ij} L/E$ and
$\Delta \tilde{m}^2_{ij} \equiv \tilde{m}^2_i - \tilde{m}^2_j$
with $\tilde{m}_i$ being the effective neutrino masses in matter. 
The relation between $\tilde{\cal J}$ and ${\cal J}$ 
reads \cite{Scott}
\begin{equation}
\tilde{\cal J} \; =\; {\cal J} ~ \frac{\Delta m^2_{21}}
{\Delta \tilde{m}^2_{21}} \cdot \frac{\Delta m^2_{31}}
{\Delta \tilde{m}^2_{31}} \cdot \frac{\Delta m^2_{32}}
{\Delta \tilde{m}^2_{32}} \;\; ,
\end{equation}
where
\begin{eqnarray}
\Delta \tilde{m}^2_{21} & = & \frac{2}{3} 
\sqrt{x^2 - 3y} \sqrt{3 (1 - z^2)} \;\; ,
\nonumber \\
\Delta \tilde{m}^2_{31} & = & \frac{1}{3}
\sqrt{x^2 - 3y} \left [ 3z + \sqrt{3 (1 - z^2)} \right ] \; ,
\nonumber \\
\Delta \tilde{m}^2_{32} & = & \frac{1}{3}
\sqrt{x^2 - 3y} \left [ 3z - \sqrt{3 (1 - z^2)} \right ] \; ,
\end{eqnarray}
and
\begin{eqnarray}
x & = & \Delta m^2_{21} + \Delta m^2_{31} + A \; , 
\nonumber \\
y & = & \Delta m^2_{21} \Delta m^2_{31} + A \left [ 
\Delta m^2_{21} \left ( 1 - |V_{e2}|^2 \right ) 
+ \Delta m^2_{31} \left ( 1 - |V_{e3}|^2 \right ) \right ] \; , 
\nonumber \\
z & = & \cos \left [ \frac{1}{3} \arccos 
\frac{2x^3 -9xy - 27 A \Delta m^2_{21} \Delta m^2_{31} |V_{e1}|^2}
{2 \left (x^2 - 3y \right )^{3/2}} \right ] \; .
\end{eqnarray}
The terrestrial matter effects are described by
the parameter $A = 2\sqrt{2} G_{\rm F} N_e E$ \cite{MSW}, with $N_e$ being
the background density of electrons and $E$ being the neutrino
beam energy. Assuming the matter density of the earth's crust to
be constant, one may get 
$A \approx 2.2 \cdot 10^{-4} ~ {\rm eV}^2 E/[{\rm GeV}]$ as a
good approximation \cite{Shrock}.

To illustrate, let us calculate $\tilde{J}$ and $\Delta \tilde{P}$
for two scenarios of the long-baseline neutrino oscillation
experiments: $L = 730$ km and $L = 2100$ km. The former baseline
corresponds to a neutrino source at Fermilab pointing toward the
Soudan mine or that at CERN toward the Gran Sasso underground 
laboratory, and the latter corresponds to a possible high-intensity
neutrino beam
from the High Energy Proton Accelerator in Tokaimura to a detector 
located in Beijing \cite{H2B}. We typically take 
$\Delta m^2_{21} \approx 5 \cdot 10^{-5} ~ {\rm eV}^2$ (the large-angle
MSW solution to the solar neutrino problem) and 
$\Delta m^2_{32} \approx 3 \cdot 10^{-3} ~ {\rm eV}^2$, as well as
$\phi_x = 90^\circ$ based on the almost bi-maximal lepton mixing pattern 
$V_{(x)}$. The numerical results of 
$\tilde{J}$ and $\Delta \tilde{P}$ as functions of the neutrino 
beam energy $E$ are shown in Figs. 1 and 2, respectively. 
We  observe that the magnitude of $\tilde{\cal J}$ can significantly 
be suppressed due to matter effects. This feature of 
$\tilde{\cal J}$ makes the measurement of leptonic CP- and T-violating
asymmetries more difficult in practice. Indeed the T-violating 
asymmetry $\Delta \tilde{P}$ is quite small in the chosen range of
the neutrino beam energy (1 GeV $\leq E \leq$ 20 GeV), at most at
the percent level. The terrestrial matter effects on $\Delta \tilde{P}$
are in general insignificant and negligible, except the case of
the resonance enhancement at $E \sim 1.5$ GeV for $L=730$ km or
at $E\sim 4$ GeV for $L =2100$ km. It should be noted that 
$\Delta \tilde{P} \approx \Delta P$ has no way to lead to 
$\tilde{\cal J} \approx {\cal J}$. Therefore a relatively clean 
signal of T violation, even measured in the future long-baseline
neutrino experiments, does not mean that the fundamental T-violating
parameter (${\cal J}$ or $\phi_x$) can {\it directly} be determined.
To pin down those genuine parameters of flavor mixing 
and T violation, we must first of all understand the terrestrial 
matter effects to a high degree of accuracy. More
reliable knowledge of the earth's matter density 
profile is unavoidably required for our long-baseline
neutrino oscillation experiments.

\section{Neutrinoless double beta decay}

So far we have only introduced a Dirac-type
T-violating phase into the lepton flavor mixing matrix $V$.
The latter may in general consist of two additional T-violating 
phases of the Majorana type; i.e.,
\begin{equation}
V \;\; \Longrightarrow \;\; \hat{V} \; =\; V P_\nu \; ,
\end{equation}
where 
$P_\nu = {\rm Diag} \{ 1, ~ e^{i\rho}, ~ e^{i\sigma} \}$
is a diagonal Majorana phase matrix. 
Although $\rho$ and $\sigma$ have no effect
on CP or T violation in normal neutrino-neutrino and 
antineutrino-antineutrino oscillations, they 
are expected to play an important role in the neutrinoless 
double beta decay, whose effective mass term is given as
\begin{equation}
\langle m_{\nu_e}\rangle \; = \; \left | \sum^3_{i=1}
\left ( m_i \hat{V}^2_{ei} \right ) \right | \; .
\end{equation}
The current experimental bound is $\langle m_{\nu_e} \rangle < 0.34$ eV,
obtained by the Heidelberg-Moscow Collaboration at the $90\%$ confidence
level \cite{HM}. For the two nearly bi-maximal lepton mixing 
scenarios under discussion, $\langle m_{\nu_e}\rangle$ reads as 
follows:
\begin{eqnarray}
\langle m_{\nu_e}\rangle_{(x)} & = & \left |\frac{\alpha}{2} c^2_x
~ + ~ \frac{\beta}{\sqrt{2}} s_x c_x e^{i\phi_x} ~ + ~
\frac{\gamma}{4} s^2_x e^{i2\phi_x} \right | \; ,
\nonumber \\
\langle m_{\nu_e}\rangle_{(z)} & = & \left |\frac{\alpha}{2} c^2_z
~ - ~ \frac{\beta}{\sqrt{2}} s_z c_z e^{i\phi_z} ~ + ~
\frac{\gamma}{4} s^2_z e^{i2\phi_z} \right | \; ,
\end{eqnarray}
where
\begin{eqnarray}
\alpha & = & m_1 ~ + ~ m_2 e^{i2\rho} \; ,
\nonumber \\
\beta & = & m_1 ~ - ~ m_2 e^{i2\rho} \; ,
\nonumber \\
\gamma & = & m_1 ~ + ~ m_2 e^{i2\rho} + 2 m_3 e^{i2\sigma} \; .
\end{eqnarray}
Note that $s_x \approx \sqrt{m_e/m_\mu} \approx 0.069$ and
$s_z \approx \sqrt{m_e/m_\tau} \approx 0.017$, therefore 
$c_x \approx c_z \approx 1$ is an excellent approximation. If the
spectrum of neutrino masses were known, one would be able to 
simplify the expression of $\langle m_{\nu_e}\rangle_{(x)}$ or
$\langle m_{\nu_e}\rangle_{(z)}$ and confront it with the present
experimental bound. Subsequently let us take four specific
but interesting cases of the neutrino mass spectrum for 
example.

(a) $m_1 \approx m_2 \approx m_3$. In this case,
the third term of $\langle m_{\nu_e}\rangle_{(x)}$ or
$\langle m_{\nu_e}\rangle_{(z)}$ is negligible. We then arrive at
\begin{eqnarray}
\langle m_{\nu_e}\rangle_{(x)} & \approx & 
m_1 \left |\frac{1}{2} \left (1 + e^{i2\rho} \right )
+ \frac{s_x}{\sqrt{2}} \left (1 - e^{i2\rho} \right ) e^{i\phi_x}
\right | \; ,
\nonumber \\
\langle m_{\nu_e}\rangle_{(z)} & \approx & 
m_1 \left |\frac{1}{2} \left (1 + e^{i2\rho} \right )
- \frac{s_z}{\sqrt{2}} \left (1 - e^{i2\rho} \right ) e^{i\phi_z}
\right | \; .
\end{eqnarray}
If $\rho \approx \pm 90^\circ$ holds, we obtain
$\langle m_{\nu_e}\rangle_{(x)} \approx \sqrt{2} ~s_x m_1$
and $\langle m_{\nu_e}\rangle_{(z)} \approx \sqrt{2} ~s_z m_1$.
The experimental bound $\langle m_{\nu_e}\rangle < 0.34$ eV is then 
assured for $m_1 \leq 3$ eV of pattern $\hat{V}_{(x)}$ or
for $m_1 \leq 15$ eV of pattern $\hat{V}_{(z)}$. If the value of
$\rho$ is not close to $\pm 90^\circ$, one may obtain
$\langle m_{\nu_e}\rangle \approx m_1 |\cos \rho|$ for both
lepton mixing patterns. Such a constraint could provide 
some information on the Majorana phase $\rho$, provided that
the magnitude of $m_1$ were already known.

(b) $m_1 \approx m_2 \gg m_3$. One may easily check that the 
results of $\langle m_{\nu_e}\rangle_{(x)}$
and $\langle m_{\nu_e}\rangle_{(z)}$ in this case 
are essentially the same as those in case (a).

(c) $m_1 \approx m_2 \ll m_3$. In this case, we obtain
$m_3 \approx \sqrt{\Delta m^2_{32}} \approx \sqrt{\Delta m^2_{\rm atm}}
\leq 0.1$ eV. Then $m_1$ and $m_2$ should be of 
${\cal O}(10^{-2})$ eV or smaller. Note that the contribution of
$m_3$ to $\langle m_{\nu_e}\rangle$ is always suppressed by
$s_x$ or $s_z$. Therefore the magnitude of $\langle m_{\nu_e}\rangle$
is at most of ${\cal O}(10^{-2})$ eV for either $\hat{V}_{(x)}$ or
$\hat{V}_{(z)}$, much smaller than the present experimental bound.

(d) $m_1 \ll m_2 \ll m_3$. This ``normal'' neutrino mass hierarchy 
\footnote{The ``inverse'' neutrino mass hierarchy $m_1 \gg m_2 \gg m_3$,
which is apparently in conflict with our choices 
$\Delta m^2_{21} \approx \pm \Delta m^2_{\rm sun}$ and
$\Delta m^2_{32} \approx \pm \Delta m^2_{\rm atm}$ 
in Eq. (1), will not be taken into account in this paper.}
leads to
$m_3 \approx \sqrt{\Delta m^2_{32}} \approx \sqrt{\Delta m^2_{\rm atm}}
\leq 0.1$ eV as well as 
$m_2 \approx \sqrt{\Delta m^2_{21}} \approx \sqrt{\Delta m^2_{\rm sun}}
\leq 0.01$ eV, where the upper limit of $m_2$ corresponds 
to the large-angle MSW solution to the
solar neutrino problem. In this case, Eq. (37) can be simplified as
\begin{eqnarray}
\langle m_{\nu_e}\rangle_{(x)} & \approx & 
\frac{1}{2} \left | m_2 e^{i2(\rho -\sigma)} 
~ + ~ m_3 s^2_x e^{i2\phi_x} \right | \; ,
\nonumber \\
\langle m_{\nu_e}\rangle_{(z)} & \approx & 
\frac{1}{2} \left | m_2 e^{i2(\rho -\sigma)} 
~ + ~ m_3 s^2_z e^{i2\phi_x} \right | \; .
\end{eqnarray}
We see that $\langle m_{\nu_e}\rangle \leq {\cal O}(10^{-2})$ eV must
hold for both nearly bi-maximal lepton mixing patterns.

The neutrinoless double beta decay itself is certainly not enough
to determine the two Majorana T-violating phases $\rho$ and $\sigma$.  
One may in principle study some other possible lepton-number-nonconserving
processes, in which the Majorana phases can show up, 
to get more constraints on $\rho$ and $\sigma$. However,
all such processes are suppressed in magnitude by an extremely small
factor compared to normal weak interactions \cite{FX01,Valle}. Hence 
it seems practically impossible to measure or constrain $\rho$ and $\sigma$
in any experiment other than the one associated with the 
neutrinoless double beta decay.

\section{Summary}

We have discussed two simple possibilities to construct the charged
lepton and neutrino mass matrices, from which two almost bi-maximal
neutrino mixing patterns can naturally emerge. Both scenarios
are favored by the atmospheric neutrino oscillation data, and are
compatible with either the large-angle (or low) MSW solution or
the vacuum oscillation solution to the solar neutrino problem. 
While the two lepton mixing patterns have
practically indistinguishable consequences on solar and atmospheric
neutrino oscillations, their predictions for leptonic CP or T violation
are different and distinguishable. Only one of them 
is likely to yield an observable T-violating asymmetry between 
$\nu_\mu \rightarrow \nu_e$ and $\nu_e \rightarrow \nu_\mu$ transitions 
in the long-baseline neutrino oscillation experiments. To be specific,
we have taken two typical baselines ($L=730$ km and $L=2100$ km) to
illustrate the magnitude of T violation and its dependence on the 
terrestrial matter effects. The implications of our nearly bi-maximal
neutrino mixing scenarios on the neutrinoless double beta decay have
also been discussed in some detail. We expect that a variety of
neutrino experiments in the near future could provide crucial tests
of the existing lepton mixing models and give useful hints towards the
symmetry or dynamics of lepton mass generation.

\vspace{0.2cm}

The author is grateful to F. Vissani for useful discussions and
enlightening comments at the early stage of this work.

\newpage

\begin{figure}[t]
\vspace{1.2cm}
\epsfig{file=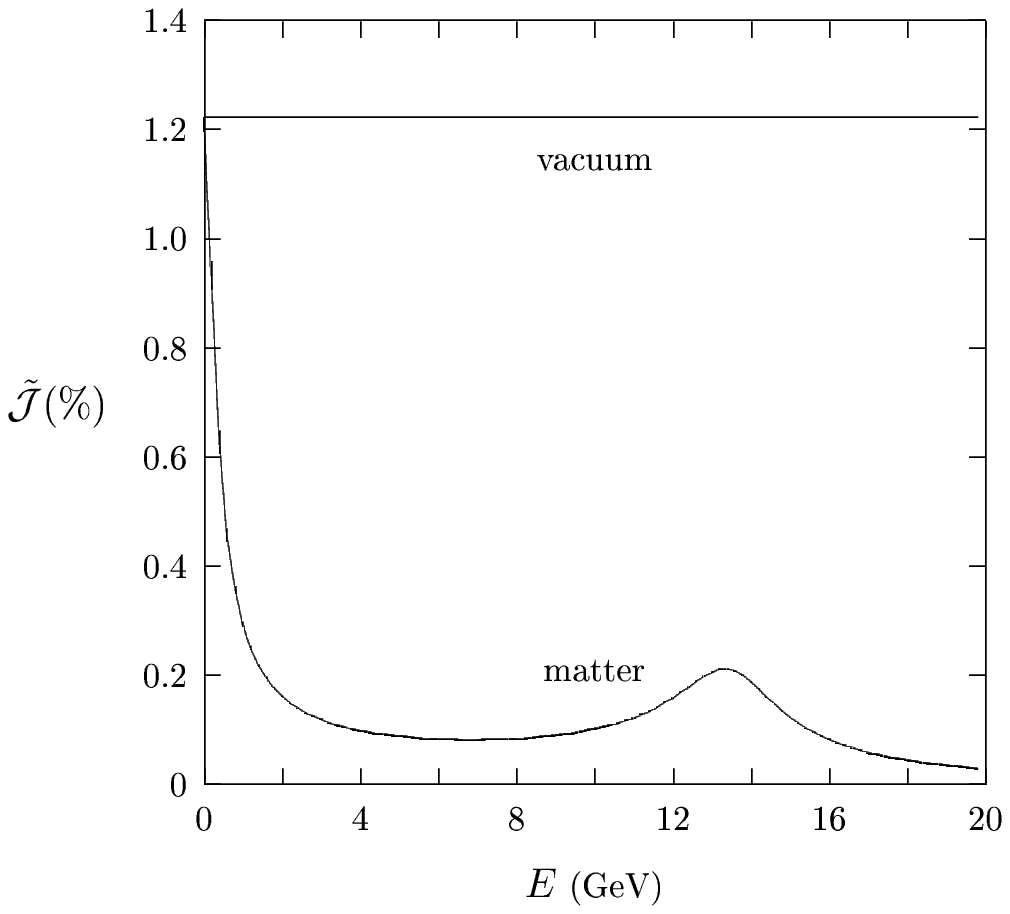,bbllx=-1cm,bblly=5.8cm,bburx=19cm,bbury=30.cm,%
width=15cm,height=20cm,angle=0,clip=}
\vspace{-9.4cm}
\caption{\small Illustrative plot for matter effects on
the universal T-violating parameter $\tilde{\cal J}$, where
$\Delta m^2_{21} \approx 5\cdot 10^{-5} ~ {\rm eV}^2$,
$\Delta m^2_{31} \approx 3\cdot 10^{-3} ~ {\rm eV}^2$,
and $\phi_x = 90^\circ$ have typically been input.}
\end{figure}

\begin{figure}[t]
\vspace{-2.2cm}
\epsfig{file=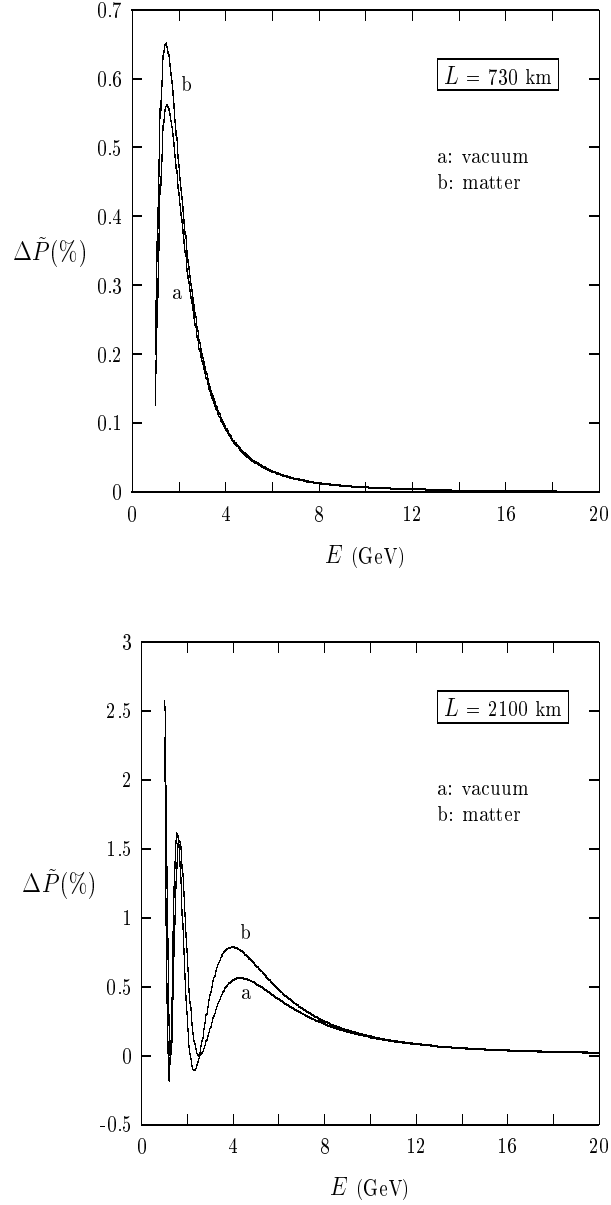,bbllx=-1cm,bblly=5.8cm,bburx=19cm,bbury=30.cm,%
width=15cm,height=20cm,angle=0,clip=}
\vspace{-1cm}
\caption{\small Illustrative plot for matter effects on
the T-violating asymmetry $\Delta \tilde{P}$ between
$\nu_\mu \rightarrow \nu_e$ and $\nu_e \rightarrow \nu_\mu$
transitions, where
$\Delta m^2_{21} \approx 5\cdot 10^{-5} ~ {\rm eV}^2$,
$\Delta m^2_{31} \approx 3\cdot 10^{-3} ~ {\rm eV}^2$,
and $\phi_x = 90^\circ$ have typically been input.}
\end{figure}

\end{document}